\documentstyle[12pt]{article}
\begin{document}
\pagestyle{empty}
\setlength{\oddsidemargin}{0.5cm}
\setlength{\evensidemargin}{0.5cm}
\setlength{\footskip}{1.5cm}
\renewcommand{\thepage}{-- \arabic{page} --}
\newcommand{\beq}{\begin{equation}}
\newcommand{\eeq}{\end{equation}}
\vspace*{-2.5cm}
\begin{flushright}
TOKUSHIMA 95-02 \\ (hep-ph/9503288) \\ March 1995
\end{flushright}
\vspace*{1.25cm}

\renewcommand{\thefootnote}{*)}
\centerline{\large\bf Remarks on the Value of the Higgs Mass}

\vskip 0.15cm
\centerline{\large\bf from the Present LEP Data}

\vspace*{1.75cm}

\centerline{\sc M. CONSOLI$^{a)}$ and Z. HIOKI$^{b)}$}

\vspace*{1.75cm}
\centerline{$a)$ Istituto Nazionale di Fisica Nucleare - Sezione di
Catania}
\centerline{\sl Corso Italia, 57 - I 95129 Catania - ITALY}

\vskip 0.3cm
\centerline{\sl $b)$ Institute of Theoretical Physics,\ University of
Tokushima}
\centerline{\sl Tokushima 770 - JAPAN}

\vspace*{3cm}

\centerline{ABSTRACT}

\vspace*{0.4cm}
\baselineskip=20pt plus 0.1pt minus 0.1pt
We perform a detailed comparison of the present LEP data with the
one-loop standard-model predictions. It is pointed out that for $m_t=
174$ GeV the ``bulk'' of the data prefers a rather large value of the
Higgs mass in the range 500-1000 GeV, in agreement with the
indications from the W mass. On the other hand, to accommodate a
light Higgs it is crucial to include the more problematic data for
the $\tau$ F-B asymmetry. We discuss further improvements on the data
taking required to obtain a firm conclusion.

\vfill
\newpage
\pagestyle{plain}
\renewcommand{\thefootnote}{\sharp\arabic{footnote}}
\setcounter{footnote}{0}
\baselineskip=21.0pt plus 0.2pt minus 0.1pt

Given the experimental evidence for the mass of the top quark $m_t=
174\pm17$ GeV from CDF \cite{CDF}, it should be possible to obtain
from the precise LEP data a precious piece of combined information on
the remaining unknown parameters of the theory, namely the Higgs mass
$m_h$ and the strong-interaction coupling constant at the Z-mass
scale $\alpha_s=\alpha_s(M_z)$. The value of $\alpha_s$, essentially
determined from the relative magnitude of the peak cross sections in
the hadronic and leptonic channels, is of primary importance both for
a test of perturbative QCD and for a comparison with Grand Unified
Theories of strong and electroweak interactions. The effect of the
Higgs boson, the remnant of Spontaneous Symmetry Breaking, on the
other hand, is rather weak since $m_h$ enters the one-loop
electroweak predictions only logarithmically and, at present, one can
only hope to separate out the heavy Higgs-mass range (say $m_h\sim$
500-1000 GeV) from the low mass regime $m_h\sim$100 GeV as predicted,
for instance, from supersymmetric theories.

In this Letter we shall present a detailed comparison between the
present LEP data, as reported by ALEPH, DELPHI, L3 and OPAL in
\cite{LEP}, and one-loop electroweak predictions for several values
of $m_h$ and $\alpha_s$. Our analysis has been essentially motivated
after exploring the range of $m_h$ associated with the measurement of
the W mass: $M_w=80.23\pm0.18$ GeV \cite{wmass}. As pointed out in
\cite{hioki}, in fact, the values of $M_w$ and $m_t$ slightly favour a
rather heavy Higgs boson even though, within their present
experimental errors, no definite conclusion is possible at present.
Therefore, it becomes important to explore the corresponding
information from LEP.\footnote{Our purpose is to study what
   information on $m_h$ we can draw from the various high-energy
   precision data. We do not consider the LR asymmetry by SLD
   \cite{SLD} since it is already known (within the standard model)
   that its present value demands a very heavy top ($>$ 200 GeV: see,
   e.g., \cite{ellis}) for the experimental bound $m_h\ >$ 61.5 GeV
   \cite{higgs}, or conversely $m_h$ must be much lower than this
   bound when taking account of $m_t$=174 GeV.}

In this context, the role of the $\alpha_s-m_h$ interdependence needs
to be discussed in some detail. In a previous analysis of the
precision LEP data (line-shape, F-B asymmetries, $\tau$-polarization,
$\cdots$), it has been stressed by Montagna et al. \cite{montagna}
that there is a non-negligible positive $m_h-\alpha_s$ correlation.
Namely, for a given top-quark mass in the range allowed by the CDF
data, the 95$\%$ CL contour includes larger values of $m_h$ when
$\alpha_s$ is allowed to vary above the Deep Inelastic Scattering
(DIS) prediction $\alpha_s(M_z)=0.113\pm0.005$ or the determination
from the hadronic shape events $\alpha_s(M_z)=0.123\pm0.006$
\cite{catani}. In this situation, constraining $\alpha_s$ in a fixed
range (say $\alpha_s=0.118\pm0.007$) may introduce uncontrolled
errors in the analysis of the purely electroweak data and, in our
opinion, should be avoided. This is even more true if one wants to
consider more ambitious theoretical frameworks since even the
simplest example of SUSY Grand Unification (with squarks and gluinos
in the 100 GeV mass range) requires $\alpha_s=0.125$ or larger
\cite{kane}. We stress that our intention is not to provide the
``best'' values of $m_h$ and $\alpha_s$. Therefore, to keep the
analysis as simple as possible, we shall first restrict to a fixed
value of the top-quark mass $m_t=174$ GeV and discuss the indications
for the Higgs mass.

To start with, we present in Table I the experimental data relevant
for our analysis. No averaging has been performed. These are the
available, individual results from the various Collaborations as
quoted in \cite{LEP} and the meaning of the various quantities is the
same as in \cite{LEP}. The theoretical predictions in Table II, for
several values of $\alpha_s$ and $m_h$ representative of the overall
situation, have been obtained with the computer code TOPAZ0 by
Montagna et al. \cite{TOPAZ0}. Finally, for the convenience of the
reader, in Tables III-VI we report the partial and total $\chi^2$ for
the various experiments and in Table VII the sum of the $\chi^2$ for
the four Collaborations.

\vskip -0.15cm
\centerline{\bf ------------------------------}
\centerline{\bf Tables I -- VII}
\centerline{\bf ------------------------------}

A few comments are in order: the global values of the $\chi^2$ in
Table VII confirm that $\alpha_s$ lies at $\sim 3\sigma$ from the DIS
prediction $\alpha_s=0.113\pm0.005$ (here, our result is in very good
agreement with the general analysis of \cite{langa} which gives
$\alpha_s=0.127\pm0.005$). Further, by inspection of Table I one
finds evidences for some systematic effect in the $\tau$ F-B
asymmetry. This effect seems to be common to all experiments and it
is confirmed from the following remark. Let us consider the global
averages reported in \cite{LEP}

$$ A^{o}_{FB}(e)=0.0156\pm0.0034~~~~~~{\rm average}   \eqno(1) $$
$$ A^{o}_{FB}(\mu)=0.0141\pm0.0021~~~~~~{\rm average} \eqno(2) $$
$$ A^{o}_{FB}(\tau)=0.0228\pm0.0026~~~~~~{\rm average}\eqno(3) $$

\noindent
and transform the averages for $A_e$ and $A_{\tau}$ \cite{LEP}

$$ A_e=0.135\pm0.011~~~~~~{ \rm average}      \eqno(4)  $$
$$ A_{\tau}=0.143\pm0.010~~~~~~{\rm average}  \eqno(5)  $$

\noindent
into ``equivalent'' F-B asymmetries by using the standard model
formula

$$ A^{o}_{FB}(1)={{3}\over{4}}A^2_e           \eqno(6)   $$
$$ A^{o}_{FB}(2)={{3}\over{4}}A_e A_{\tau}    \eqno(7)   $$
we find
$$ A^{o}_{FB}(1)=0.0137\pm0.0023     \eqno(8)    $$
$$ A^{o}_{FB}(2)=0.0145\pm0.0023     \eqno(9)    $$
in very good agreement with Eqs.(1,2) but not with Eq.(3). Therefore,
there may be some problem in the direct measurement of
$A^{o}_{FB}(\tau)$ since all other measurements are in excellent
agreement with each other. Just to have an idea of the effect, if the
data for the $\tau$ F-B asymmetry are omitted in the evaluation of
the $\chi^2$ we find the results illustrated in Table VIII which one
should compare with Tables III- VII. The ``bulk'' of the LEP data,
namely those well consistent with each other, show no preference for
a light Higgs boson and the best values of the $\chi^2$ are obtained
for a large value of $m_h$, just as in the case of the W mass
reported in \cite{hioki}.

\vskip -0.15cm
\centerline{\bf ------------------------------}
\centerline{\bf Table VIII}
\centerline{\bf ------------------------------}

Finally, to have an idea of the dependence on $m_t$, we report in
Table IX and Table X the total $\chi^2$ for $m_t$=160, 174 and 190
GeV including all data or excluding $A^{o}_{FB}(\tau)$. As first
noticed by Ellis et al. \cite{ellis,fogli}, by increasing
(decreasing) the top-quark mass a larger (smaller) value of $m_h$ is
favoured and the shape of the $\chi^2$ is well consistent with all
values of the Higgs mass. For $m_t=174$ GeV, however, Table IX and
Table X give rather different information and it becomes crucial to
include the more problematic data for $A^{o}_{FB}(\tau)$ to
accommodate values $m_h\sim$ 100 GeV.

\vskip -0.15cm
\centerline{\bf ------------------------------}
\centerline{\bf Tables IX and X}
\centerline{\bf ------------------------------}

Conclusions: we do not know whether Tables VIII and X represent a
more faithful representation of the real physical situation than
Tables VII and IX. Most likely, our results suggest only that further
improvement in the data taking is needed for a definitive answer. Our
analysis confirms that the possibility to obtain precious information
on the Higgs mass is not unrealistic if the top-quark mass is
measured with a higher precision at the Tevatron. However, {\it if}
we really want to explore the full potentiality of LEP for a precise
determination of $m_h$ in the standard electroweak theory, many
systematic effects have still to be understood and much more
stringent tests have to be performed. To this end, a precise scanning
of the Z resonance with 4 or more points at high statistics off peak
cannot be postponed anymore ( $\sim 90\%$ of the total events have
been collected at the pole ). Further, a high luminosity phase of LEP
I, where each collaboration will detect millions of Z's per run and
the purely statistical errors will become negligible, is needed to
obtain a definitive consistency check of the systematics of the
various experiments.

\vskip 20pt
\centerline{ACKNOWLEDGEMENTS}
\par We thank Giampiero Passarino for many useful discussions.

\vskip 0.8cm
\centerline{\bf Note added in proof}
\par After completing this work, new data for the top mass have been
published by the D0 and CDF Collaborations \cite{new}. Their
results, $m_t=199^{+19}_{-21}(\rm stat.)\pm 22(\rm syst.)$ GeV from
D0 and $m_t=176\pm 8(\rm stat.) \pm 10(\rm syst.) $ GeV from CDF,
favour a slightly higher ($m_t=180\pm 12$ GeV) than the one mainly
used in our analysis and enforce the preference for a rather heavy
Higgs particle deduced from the ``bulk'' of the LEP data. Tables IX
and X become as follows if we add a term $((m_t -180)/12)^2$ to the
$\chi^2$ computation:
\begin{center}
{\bf (Table IX)}

\vspace{0.5cm}
\begin{tabular}{ccccc}
   $\alpha_s$      &~~$0.113$~~&~~$0.125$~~&~~$0.127$~~&~~$0.130$~~\\
   $m_h$(GeV)      &  $100$    &  $100$    &  $500$    &  $1000$   \\
\hline\hline
 $m_t$(GeV) =160~~~&  $48.4$   &  $38.7$   &  $42.9$   &  $46.9$   \\
\hline
\phantom{
 $m_t$(GeV)}=174~~~&  $43.9$   &  $38.1$   &  $36.7$   &  $38.5$   \\
\hline
\phantom{
 $m_t$(GeV)}=190~~~&  $47.6$   &  $45.6$   &  $37.9$   &  $36.6$   \\
\hline
\end{tabular}

\vspace{1cm}
{\bf (Table X)}

\vspace{0.5cm}
\begin{tabular}{ccccc}
   $\alpha_s$      &~~$0.113$~~&~~$0.125$~~&~~$0.127$~~&~~$0.130$~~\\
   $m_h$(GeV)      &  $100$    &  $100$    &  $500$    &  $1000$   \\
\hline\hline
 $m_t$(GeV) =160~~~&  $39.1$   &  $29.2$   &  $29.8$   &  $31.6$   \\
\hline
\phantom{
 $m_t$(GeV)}=174~~~&  $36.3$   &  $30.3$   &  $25.7$   &  $25.8$   \\
\hline
\phantom{
 $m_t$(GeV)}=190~~~&  $41.8$   &  $39.8$   &  $29.3$   &  $26.6$   \\
\hline
\end{tabular}
\end{center}


\newpage
\renewcommand{\arraystretch}{1.4}
\centerline{\large TABLES}

\vskip 2cm
\begin{center}
\begin{tabular}{lcccc}
\ ~~ &~~~ {ALEPH} &~~~ {DELPHI} &~~~ {L3} &~~~ {OPAL} \\ \hline\hline
$\Gamma_z(MeV)$ &~~~$2495.9\pm6.1$ & ~~~$2495.1\pm5.9$ &
\ ~$2504.0\pm5.8$ &~~~$2494.5\pm6.1$ \\ \hline
$\sigma_{had}$(nb) &~~~$41.59\pm0.13$ & ~~~$41.26\pm0.17$&
\ ~$41.44\pm0.15$&~~~$41.47\pm0.16$ \\ \hline
$R_e$ &~~~$20.67\pm0.13$ & ~~~$20.96\pm0.16$ &~~~$20.94\pm0.13$ &
\ ~$20.90\pm0.13$ \\ \hline
$R_{\mu}$ &~~~$20.91\pm0.14$ & ~~~$20.60\pm0.12$ &
\ ~$20.93\pm0.14$ &~~~$20.855\pm0.097$ \\ \hline
$R_{\tau}$ &~~~$20.69\pm0.12$ & ~~~$20.64\pm0.16$ &
\ ~$20.70\pm0.17$ &~~~$20.91\pm0.13$ \\ \hline
$A^{o}_{FB}(e)$ &~~~$0.0212\pm0.0054$ & ~~~$0.0207\pm0.0073$ &
\ ~$0.0109\pm0.0081$ &~~~$0.0060\pm0.0066$ \\ \hline
$A^{o}_{FB}(\mu)$ &~~~$0.0189\pm0.0039$ & ~~~$0.0128\pm0.0037$ &
\ ~$0.0132\pm0.0048$ &~~~$0.0124\pm0.0035$ \\ \hline
$A^{o}_{FB}(\tau)$ &~~~$0.0253\pm0.0043$ & ~~~$0.0209\pm0.0057$ &
\ ~$0.0299\pm0.0073$ &~~~$0.0193\pm0.0044$ \\ \hline
$A_e$ &~~~$0.127\pm0.017$ & ~~~$0.140\pm0.028$ &
\ ~$0.154\pm0.023$ &~~~$0.122\pm0.032$ \\ \hline
$A_{\tau}$ &~~~$0.137\pm0.014$ & ~~~$0.144\pm0.024$ &
\ ~$0.144\pm0.020$ &~~~$0.153\pm0.023$ \\ \hline
\end{tabular}
\end{center}

\vskip 1cm \noindent
{\bf Table I.}\ The experimental data from the four LEP
Collaborations.
\newpage
\vspace*{1.5cm}
\begin{center}
\begin{tabular}{lcccc}
\ $\alpha_s$ &~~~~ {$0.113$} &~~~~ {$0.125$} &~~~~ {$0.127$} &
\ ~~ {$0.130$} \\
\ $m_h$(GeV) &~~~~ {$100$} &~~~~ {$100$} &~~~~ {$500$} &
\ ~~ {$1000$} \\ \hline\hline
$\Gamma_z$(MeV) &~~~~$2494.2$ & ~~~~$2500.7$ &~~~~$2496.5$ &
\ ~~$2495.1$ \\ \hline
$\sigma_{had}$(nb) &~~~~$41.506$ & ~~~~$41.444$&~~~~$41.434$&
\ ~~$41.421$ \\ \hline
$R_e$ &~~~~$20.706$ & ~~~~$20.786$ &~~~~$20.787$ &~~~~$20.798$ \\
\hline
$R_{\mu}$ &~~~~$20.706$ & ~~~~$20.786$ &~~~~$20.787$ &~~~~$20.798$ \\
\hline
$R_{\tau}$ &~~~~$20.753$ & ~~~~$20.834$ &~~~~$20.835$ &
\ ~~$20.846$ \\ \hline
$A^{o}_{FB}(e)$ &~~~~$0.0170$ & ~~~~$0.0169$ &~~~~$0.0154$ &
\ ~~$0.0147$ \\ \hline
$A^{o}_{FB}(\mu)$ &~~~~$0.0170$ & ~~~~$0.0169$ &~~~~$0.0154$ &
\ ~~$0.0147$ \\ \hline
$A^{o}_{FB}(\tau)$ &~~~~$0.0170$ & ~~~~$0.0169$ &~~~~$0.0154$ &
\ ~~$0.0147$ \\ \hline
$A_e$ &~~~~$0.1505$ & ~~~~$0.1502$ &~~~~$0.1433$ &~~~~$0.1400$ \\
\hline
$A_{\tau}$ &~~~~$0.1505$ & ~~~~$0.1502$ &~~~~$0.1433$ &~~~~$0.1400$\\
\hline
\end{tabular}
\end{center}

\vskip 1cm \noindent
{\bf Table II.}\ We report the theoretical predictions at various
values of $\alpha_s(M_z)$ and $m_h$ for a fixed top-quark mass
$m_t=174$ GeV. These predictions have been obtained with the computer
code TOPAZ0 by G. Montagna, O. Nicrosini, G. Passarino and
F. Piccinini.
\newpage
\vspace*{1.5cm}
\begin{center}
ALEPH

\vskip 30 pt
\begin{tabular}{lcccc}
\ $\alpha_s$ &~~~~ {$0.113$} &~~~~ {$0.125$} &~~~~ {$0.127$} &
\ ~~ {$0.130$} \\
\ $m_h$(GeV) &~~~~ {$100$} &~~~~ {$100$} &~~~~ {$500$} &
\ ~~ {$1000$} \\ \hline\hline
$\Gamma_z$ &~~~~$0.08$ & ~~~~$0.62$ &~~~~$0.01$ &~~~~$0.02$ \\ \hline
$\sigma_{had}$ &~~~~$0.42$ & ~~~~$1.26$&~~~~$1.44$&~~~~$1.69$ \\
\hline
$R_e$ &~~~~$0.08$ & ~~~~$0.80$ &~~~~$0.81$ &~~~~$0.97$ \\ \hline
$R_{\mu}$ &~~~~$2.12$ & ~~~~$0.78$ &~~~~$0.77$ &~~~~$0.64$ \\ \hline
$R_{\tau}$ &~~~~$0.28$ & ~~~~$1.44$ &~~~~$1.46$ &~~~~$1.69$ \\ \hline
$A^{o}_{FB}(e)$ &~~~~$0.60$ & ~~~~$0.63$ &~~~~$1.15$ &~~~~$1.45$ \\
\hline
$A^{o}_{FB}(\mu)$ &~~~~$0.24$ & ~~~~$0.26$ &~~~~$0.81$ &~~~~$1.16$ \\
\hline
$A^{o}_{FB}(\tau)$ &~~~~$3.73$ & ~~~~$3.82$ &~~~~$5.30$ &~~~~$6.08$\\
\hline
$A_e$ &~~~~$1.91$ & ~~~~$1.86$ &~~~~$0.92$ &~~~~$0.58$ \\ \hline
$A_{\tau}$ &~~~~$0.93$ & ~~~~$0.89$ &~~~~$0.20$ &~~~~$0.05$ \\
\hline\hline
TOTAL $\chi^2$ &~~~~$10.4$ & ~~~~$12.4$ &~~~~$12.9$ &~~~~$14.3$ \\
\hline
\end{tabular}
\end{center}

\vskip 1cm \noindent
{\bf Table III.}\ Individual and total $\chi^2$ from the ALEPH
Collaboration at various values of $\alpha_s(M_z)$ and $m_h$ for
$m_t=$174 GeV.
\newpage
\vspace*{1.5cm}
\begin{center}
DELPHI

\vskip 30 pt
\begin{tabular}{lcccc}
\ $\alpha_s$ &~~~~ {$0.113$} &~~~~ {$0.125$} &~~~~ {$0.127$} &
\ ~~ {$0.130$} \\
\ $m_h$(GeV) &~~~~ {$100$} &~~~~ {$100$} &~~~~ {$500$} &
\ ~~ {$1000$} \\ \hline\hline
$\Gamma_z$ &~~~~$0.02$ & ~~~~$0.90$ &~~~~$0.06$ &~~~~$0.00$ \\ \hline
$\sigma_{had}$ &~~~~$2.09$ & ~~~~$1.17$&~~~~$1.05$&~~~~$0.90$ \\
\hline
$R_e$ &~~~~$2.52$ & ~~~~$1.18$ &~~~~$1.17$ &~~~~$1.03$ \\ \hline
$R_{\mu}$ &~~~~$0.78$ & ~~~~$2.40$ &~~~~$2.43$ &~~~~$2.72$ \\ \hline
$R_{\tau}$ &~~~~$0.50$ & ~~~~$1.47$ &~~~~$1.49$ &~~~~$1.66$ \\ \hline
$A^{o}_{FB}(e)$ &~~~~$0.26$ & ~~~~$0.27$ &~~~~$0.53$ &~~~~$0.68$ \\
\hline
$A^{o}_{FB}(\mu)$ &~~~~$1.29$ & ~~~~$1.23$ &~~~~$0.49$ &~~~~$0.26$ \\
\hline
$A^{o}_{FB}(\tau)$ &~~~~$0.47$ & ~~~~$0.49$ &~~~~$0.93$ &~~~~$1.18$\\
\hline
$A_e$ &~~~~$0.14$ & ~~~~$0.13$ &~~~~$0.01$ &~~~~$0.00$ \\ \hline
$A_{\tau}$ &~~~~$0.07$ & ~~~~$0.07$ &~~~~$0.00$ &~~~~$0.03$ \\
\hline\hline
TOTAL $\chi^2$ &~~~~$8.1$ & ~~~~$9.3$ &~~~~$8.2$ &~~~~$8.5$ \\ \hline
\end{tabular}
\end{center}

\vskip 1cm \noindent
{\bf Table IV.}\ The same as in Table III for the DELPHI
Collaboration.
\newpage
\vspace*{1.5cm}
\begin{center}
L3

\vskip 30 pt
\begin{tabular}{lcccc}
\ $\alpha_s$ &~~~~ {$0.113$} &~~~~ {$0.125$} &~~~~ {$0.127$} &
\ ~~ {$0.130$} \\
\ $m_h$(GeV) &~~~~ {$100$} &~~~~ {$100$} &~~~~ {$500$} &
\ ~~ {$1000$} \\ \hline\hline
$\Gamma_z$ &~~~~$2.85$ & ~~~~$0.32$ &~~~~$1.67$ &~~~~$2.35$ \\ \hline
$\sigma_{had}$ &~~~~$0.19$ & ~~~~$0.00$&~~~~$0.00$&~~~~$0.02$ \\
\hline
$R_e$ &~~~~$3.24$ & ~~~~$1.40$ &~~~~$1.39$ &~~~~$1.19$ \\ \hline
$R_{\mu}$ &~~~~$2.56$ & ~~~~$1.06$ &~~~~$1.04$ &~~~~$0.89$ \\ \hline
$R_{\tau}$ &~~~~$0.10$ & ~~~~$0.62$ &~~~~$0.63$ &~~~~$0.74$ \\ \hline
$A^{o}_{FB}(e)$ &~~~~$0.57$ & ~~~~$0.55$ &~~~~$0.31$ &~~~~$0.22$ \\
\hline
$A^{o}_{FB}(\mu)$ &~~~~$0.63$ & ~~~~$0.63$ &~~~~$0.21$ &~~~~$0.10$ \\
\hline
$A^{o}_{FB}(\tau)$ &~~~~$3.12$ & ~~~~$3.17$ &~~~~$3.95$ &~~~~$4.33$\\
\hline
$A_e$ &~~~~$0.02$ & ~~~~$0.03$ &~~~~$0.22$ &~~~~$0.37$ \\ \hline
$A_{\tau}$ &~~~~$0.11$ & ~~~~$0.10$ &~~~~$0.00$ &~~~~$0.04$ \\
\hline\hline
TOTAL $\chi^2$ &~~~~$13.4$ & ~~~~$7.9$ &~~~~$9.4$ &~~~~$10.2$ \\
\hline
\end{tabular}
\end{center}

\vskip 1cm \noindent
{\bf Table V.}\ The same as in Table III for the L3 Collaboration.
\newpage
\vspace*{1.5cm}
\begin{center}
OPAL

\vskip 30 pt
\begin{tabular}{lcccc}
\ $\alpha_s$ &~~~~ {$0.113$} &~~~~ {$0.125$} &~~~~ {$0.127$} &
\ ~~ {$0.130$} \\
\ $m_h$(GeV) &~~~~ {$100$} &~~~~ {$100$} &~~~~ {$500$} &
\ ~~ {$1000$} \\ \hline\hline
$\Gamma_z$ &~~~~$0.05$ & ~~~~$1.03$ &~~~~$0.11$ &~~~~$0.01$ \\ \hline
$\sigma_{had}$ &~~~~$0.05$ & ~~~~$0.03$&~~~~$0.05$&~~~~$0.09$ \\
\hline
$R_e$ &~~~~$2.23$ & ~~~~$0.77$ &~~~~$0.76$ &~~~~$0.62$ \\ \hline
$R_{\mu}$ &~~~~$2.36$ & ~~~~$0.51$ &~~~~$0.49$ &~~~~$0.34$ \\ \hline
$R_{\tau}$ &~~~~$1.46$ & ~~~~$0.34$ &~~~~$0.33$ &~~~~$0.27$ \\ \hline
$A^{o}_{FB}(e)$ &~~~~$2.78$ & ~~~~$2.73$ &~~~~$2.03$ &~~~~$1.74$ \\
\hline
$A^{o}_{FB}(\mu)$ &~~~~$1.73$ & ~~~~$1.65$ &~~~~$0.73$ & ~~~$0.43$ \\
\hline
$A^{o}_{FB}(\tau)$ &~~~~$0.27$ & ~~~~$0.30$ &~~~~$0.79$ & ~~~$1.09$
\\ \hline
$A_e$ &~~~~$0.79$ & ~~~~$0.78$ &~~~~$0.44$ &~~~~$0.32$ \\ \hline
$A_{\tau}$ &~~~~$0.01$ & ~~~~$0.01$ &~~~~$0.18$ &~~~~$0.32$ \\
\hline\hline
TOTAL $\chi^2$ &~~~~$11.7$ & ~~~~$8.2$ &~~~~$5.9$ &~~~~$5.2$ \\
\hline
\end{tabular}
\end{center}

\vskip 1cm \noindent
{\bf Table VI.}\ The same as in Table III for the OPAL Collaboration.
\newpage
\vspace*{-1cm}
\begin{center}
ALEPH+DELPHI+L3+OPAL

\vskip 30 pt
\begin{tabular}{lcccc}
\ $\alpha_s$ &~~~~ {$0.113$} &~~~~ {$0.125$} &~~~~ {$0.127$} &
\ ~~ {$0.130$} \\
\ $m_h$(GeV) &~~~~ {$100$} &~~~~ {$100$} &~~~~ {$500$} &
\ ~~ {$1000$} \\ \hline
TOTAL $\chi^2$ &~~~~$43.6$ & ~~~~$37.8$ &~~~~$36.4$ &~~~~$38.2$ \\
\hline
\end{tabular}
\end{center}

\vskip 1cm \noindent
{\bf Table VII.}\ Total $\chi^2$ for the four Collaborations.

\vspace{2.5cm}
\begin{center}
\begin{tabular}{lcccc}
\ $\alpha_s$ &~~~~ {$0.113$} &~~~~ {$0.125$} &~~~~ {$0.127$} &
\ ~~ {$0.130$} \\
\ $m_h$(GeV) &~~~~ {$100$} &~~~~ {$100$} &~~~~ {$500$} &
\ ~~ {$1000$} \\ \hline\hline
ALEPH  &~~~~$6.7$  & ~~~~$8.6$ &~~~~$7.6$ &~~~~$8.2$ \\ \hline
DELPHI &~~~~$7.6$  & ~~~~$8.8$ &~~~~$7.3$ &~~~~$7.3$ \\ \hline
L3     &~~~~$10.3$ & ~~~~$4.7$ &~~~~$5.4$ &~~~~$5.9$ \\ \hline
OPAL   &~~~~$11.4$ & ~~~~$7.9$ &~~~~$5.1$ &~~~~$4.1$ \\ \hline\hline
TOTAL $\chi^2$ &~~~~$36.0$ & ~~~~$30.0$ &~~~~$25.4$ &~~~~$25.5$ \\
\hline
\end{tabular}
\end{center}

\vskip 1cm \noindent
{\bf Table VIII.}\ Total $\chi^2$ for the four Collaborations by
excluding the data for $A^{o}_{FB}(\tau)$.
\newpage
\vspace*{-1cm}
\begin{center}
ALEPH+DELPHI+L3+OPAL

\vskip 30 pt
\begin{tabular}{lccccc}
$m_t$(GeV)&~~~$\alpha_s$ &~~~~ {$0.113$} &~~~~ {$0.125$} &
\ ~~ {$0.127$} &~~~~ {$0.130$} \\
\ ~~~~&~~~$m_h$(GeV) &~~~~ {$100$} &~~~~ {$100$} &~~~~ {$500$} &
\ ~~ {$1000$} \\ \hline\hline
160 &~~~~~~&~~~~$45.6$ & ~~~~$35.9$ &~~~~$40.1$ &~~~~$44.1$ \\ \hline
174 &~~~~~~&~~~~$43.6$ & ~~~~$37.8$ &~~~~$36.4$ &~~~~$38.2$ \\ \hline
190 &~~~~~~&~~~~$46.9$ & ~~~~$44.9$ &~~~~$37.2$ &~~~~$35.9$ \\ \hline
\end{tabular}
\end{center}

\vskip 1cm \noindent
{\bf Table IX.}\ Total $\chi^2$ for the four Collaborations at
various values of $m_t$.

\vspace{2cm}
\begin{center}
ALEPH+DELPHI+L3+OPAL

\vskip 30 pt
\begin{tabular}{lccccc}
$m_t$(GeV)&~~~$\alpha_s$ &~~~~ {$0.113$} &~~~~ {$0.125$} &
\ ~~ {$0.127$} &~~~~ {$0.130$} \\
\ ~~~~~&~~~$m_h$(GeV) &~~~~ {$100$} &~~~~ {$100$} &~~~~ {$500$} &
\ ~~ {$1000$} \\ \hline\hline
160 &~~~~~~~&~~~~$36.3$ & ~~~~$26.4$ &~~~~$27.0$ &~~~~$28.8$ \\
\hline
174 &~~~~~~~&~~~~$36.0$ & ~~~~$30.0$ &~~~~$25.4$ &~~~~$25.5$ \\
\hline
190 &~~~~~~~&~~~~$41.1$ & ~~~~$39.1$ &~~~~$28.6$ &~~~~$25.9$ \\
\hline
\end{tabular}
\end{center}

\vskip 1cm \noindent
{\bf Table X.}\ Total $\chi^2$ for the four Collaborations at various
values of $m_t$ by excluding the data for $A^{o}_{FB}(\tau)$.

\newpage
\begin{thebibliography}{99}
\bibitem{CDF} CDF Collaboration : F. Abe et al., Phys. Rev. Lett.
{\bf 73} (1994), 225.
%
\bibitem{LEP} The LEP Collaborations, {\it Combined preliminary data
on Z parameters from the LEP experiments and constraints on the
Standard Model}, preprint CERN/PPE/94-187, 1994.
%
\bibitem{wmass} M. Demarteu, H. Fisch, U. Heintz, R. Keup and
D. Saltzberg, Joint CDF note/D0 note, CDF/PHYS/CDF/PUBLIC/2552 and
D0NOTE 2115.
%
\bibitem{hioki} Z. Hioki and R. Najima, Mod. Phys. Lett. {\bf A10}
(1995), 121.
%
\bibitem{SLD} SLD Collaboration : K. Abe et al., {\sl Phys. Rev.
Letters} {\bf 73} (1994), 25.
%
\bibitem{ellis} J. Ellis, G. L. Fogli and E. Lisi, Phys. Lett.
{\bf B333} (1994) 118.
%
\bibitem{higgs} T. Kobayashi, Preprint UT-ICEPP 94-05 (Talk at the
22th INS International Symposium on Physics with High Energy
Colliders, Tokyo, Japan, March 8-10, 1994)
%
\bibitem{montagna} G. Montagna, O. Nicrosini, G. Passarino and
F. Piccinini, Phys. Lett. {\bf B355} (1994), 484.
%
\bibitem{catani} S. Catani, {\it QCD and Jets (Theory)}, Plenary Talk
at the International Europhysics Conference on High Energy Physics,
Marseille, 22-28 July 1993, preprint DFF 194/11/93, November 1993.
%
\bibitem{kane} G. Kane, C. Kolda, L. Roszkowski and J. Wells,
Phys. Rev. {\bf D49} (1994), 6173.
%
\bibitem{TOPAZ0} G. Montagna, O. Nicrosini, G. Passarino,
F. Piccinini and R. Pittau, Comput. Phys. Commun. {\bf 76} (1993),
328.
%
\bibitem{langa} P. Langacker, {\it Tests of the Standard Model and
Searches for New Physics}, to be published in {\it Precisions Tests
of the Standard Electroweak Model}, World Scientific, Singapore 1994
[hep-ph/9412361].
%
\bibitem{fogli} J. Ellis, G. L. Fogli and E. Lisi, Phys. Lett.
{\bf B318} (1993) 148 and older references quoted therein.
%
\bibitem{new} CDF Collaboration : F. Abe et al.,
preprint FERMILAB-PUB-95/022-E;\\
D0 Collaboration : S. Abachi et al., preprint FERMILAB-PUB-95/028-E.
\end{thebibliography}
\end{document}